\documentstyle[psfig,supertab]{article}
\begin{document}

{\large

\title{Optical spectrum  of the IR-source IRC+10420  in 1992-1996}

\author{Klochkova V.G., Chentsov E.L., Panchuk V.E.}

\maketitle

\centerline{\it Special Astrophysical Observatory, Nizhnij Arkhyz, 357147 RUSSIA}

\begin{abstract}
To understand the evolutionary stage of the peculiar supergiant
IRC+10420,  we have been taking spectra for several years at the 6\,m telescope.
The optical spectrum of IRC+10420 of the years from 1992 through 1996
points to the increase in the tempe\-ra\-ture: spectral class A5 instead of
the former F8, as was pointed out by Humpreys et al., (1973).
Now it resembles the spectra of late-type B[e] stars.
The spectrum contains absorptions (mainly of ions) formed in the photosphere,
apparently stationary with respect to the star center of mass, and emissions
too, which can be formed in the fossil  expanding envelope as well as partly in
its compressing region.

Using our spectra and spectral data obtained by Oudmaijer (1995) we estimated
the atmospheric parameters ${\rm T_{eff} = 8500\,K}$, log\,g=1.0,
${{\rm \xi_t = 12\,km/s}}$ and concluded that metallicity of IRC+10420 is solar:
the average value ${\rm [(V,Cr,Fe)/H]_{\odot} = -0.03}$.

Combination of results allows us to consider IRC+10420 as a massive supergiant
evolving to the WR-stage.
\end{abstract}


\section{Introduction}

The OH/IR source IRC+10420  = IRAS19244+1115 identified with the peculiar high
luminosity star V1302\,Aql is a unique object, which has been carefully
and comprehensively studied over the last
decades but still remains a puzzle. Of the two hypotheses about its nature
neither seems to be convincingly preponderant  as yet. According to Fix and
Cobb (1987), Hrivnak et al., (1989) and others this is a degenerate core giant
evolving through the proto-planetary nebula stage with a luminosity no higher
than ${\rm 5 * 10^{4} L_{\odot}}$. According to Jones et al., (1993), Humphreys
and Davidson (1994) and Oudmaijer et al. (1996) this is a core-burning
hypergiant of ${\rm \approx 5 * 10^{5} L_{\odot}}$.

The difficulty of choice is due to:
\begin{itemize}
\item{the uncertainty of fundamental observational parameters, such as spectral
      class and distance;}
\item{the fact that with the difference in mass, age, even type of stellar
population the evolutionary processes and their observational evidence  are
similar: in both alternatives the effective temperature of the star increases,
there is a gaseous-dust envelope inherited from the red giant or supergiant
phase, which interacts with the stellar wind;}
\item{the presence of several
competing models: thin chromosphere in the expanding gaseous-dust envelope
such optically thick that we see the light of the star being multiple
scaterring by circumstellar dust (Fix and Cobb, 1987);
a gaseous-dust disk in a clumpy envelope (Jones et al., 1993); jets
with a small angle of opening  (Oudmaijer et al., 1994); and at last infall
of circumstellar material onto photosphere (Oudmaijer, 1995).}
\end{itemize}
Outer regions of the source of radius 2-3 arcsec are mainly described and mapped
by radio astronomy techniques from emission of molecules (paper by Nedoluha and
Bowers (1992) and references therein). The gas envelope has been imaged by the
methods of infrared speckle interferometry  and coronagraphy (Ridgway et al.,
1986OC; Kastner and Weintraub, 1995).  The image of the dust envelope also
extends to several seconds of arc, but the radiation is sharply enhanced in
the region of about 0.1\,arcsec in radius. However space  or ground\--based speckle
images in the optical range, which could refine the structure of the central
region, unknown to us.

Here the high resolution optical spectrum of IRC+10420, which
contains the emission and absorption lines of the envelope and absorption ones
of the photosphere or of the pseudophotosphere of the central star, is described.

\begin{table}
\caption{Observation log of IRC+10420}
\begin{tabular}{ccc}
&&\\[5pt]
\hline
  Spectrum & Date & Spectral interval, ${\rm \AA}$\\
\hline
S03503 & 21.08.92. & 5500-7200  \\
S13003 & 11.12.95. & 4800-6700 \\
S13714 & 01.05.96. & 5200-6800 \\
S14409 & 03.07.96. & 5200-7900 \\
\hline
\end{tabular}
\end{table}

\section{Observational data}

 Spectra of IRC+10420 were taken with the CCD equipped echelle
spectrometer LYNX mounted at the Nasmyth focus of the 6 m telescope of
SAO RAS (Panchuk et al., 1993). The log of our observations is given in Table\,1.

The average spectral resolution  was 0.3 m\AA. The signal-to-noise ratio was
equal to 20-50 for different  spectral orders. Before 1995 a CCD of 530\,x\,580
pixels was used and the low echelle orders did not overlap, in the two latter
spectra a CCD of 1140\,x\,1170 pixels was employed, and the indicated region
overlapped completely. The MIDAS system was used for echelle-images reduction.
The comparison spectrum source was an argon-filled thorium hollow-cathode lamp.
Control and correction of instrumental displacement of the comparison and
object spectra were done  by the telluric lines ${\rm O_{2}}$ and ${\rm
H_{2}O}$. The final accidental radial velocity determination errors are
listed in Table\,2. Systematic errors may reach 1-2\, km/s.

\begin{table}
\caption{Heliocentric radial velocities in km/s for groups of lines
               in the spectra of IRC+10420 in 1995-1996}
\begin{tabular}{lrr@{}lr@{}l}
&&&&& \\
\hline
Group of lines & n &\multicolumn{4}{c}{Vr}\\
\cline{3-6}
&&\multicolumn{2}{c}{{\rm em}} & \multicolumn{2}{c}{{\rm abs}} \\
\hline
${\rm [OI], [CaII],[FeII]}$ & 6 &        & +66 $\pm $ 2  &            &            \\
          NI(3)            & 3 &        &                &    & +66 $\pm $ 2                 \\
          FeII(40,46)      & 7 &        & +60 $\pm $ 3   &        &              \\
         SiII(2)           & 2 &        &                &        & +72 $\pm $ 2 \\
        ScII,TiII,CrII     & 6 &        &                & $\ge$  & +82 $\pm $ 2 \\
        FeII(48,49,74)     &12 & $\le $ &+40 $\pm $ 3    & $\ge$  & +91 $\pm $3  \\
 ${\rm H_{\alpha, \beta}}$ & 2 & $\le $ & +15, +132      & $\ge $ & +71 $\pm $ 2 \\
          NaI(1), KI(1)    & 4 &        &                &        & -5: +25:+(83-92) \\
             DIB           &11 &        &                & &-3 $\pm $ 2 \\
\hline
\end{tabular}
\end{table}

\section{Results and discussion}

The only difference between the IRC+10420 spectra of 1992 and 1995-1996 we
have noticed is the enhanced emission lines. Some emission lines, in particular
close to $\lambda\lambda $ 6963, 7065, 7382, 7502, 7722\,${\rm \AA\AA}$,
distinctly shown in our last spectrum S14409 were absent both in our 1992
spectrum and in Oudmaijer's spectrum which was obtained in 1994.

The spectral lines are united in groups by likeness of the profiles and closeness
of the radial velocity values.
The averaged profiles derived from the spectra  S13714 and S14409 of higher
quality for individual typical lines are shown in Fig.\,1. The heliocentric
radial velocities for the emission and absorption components of the profiles
separately, also averaged over the groups of lines, are given in Table\,2. In
the second column of  Table\,2 are presented the numbers of lines in the
groups, which were used for radial velocity measurements.
As it is seen from Fig.\,1, the line profiles are very different - from
slightly asymmetric emissions details to undistorted absorptions lines
through more complex intermidate forms.
The variations in the shape of our line profiles follow the variations
in the optical and IR lines according to Oudmaijer et al. (1994),
and Oudmaijer (1995).

\begin{figure}
\par
\centerline{\psfig{figure=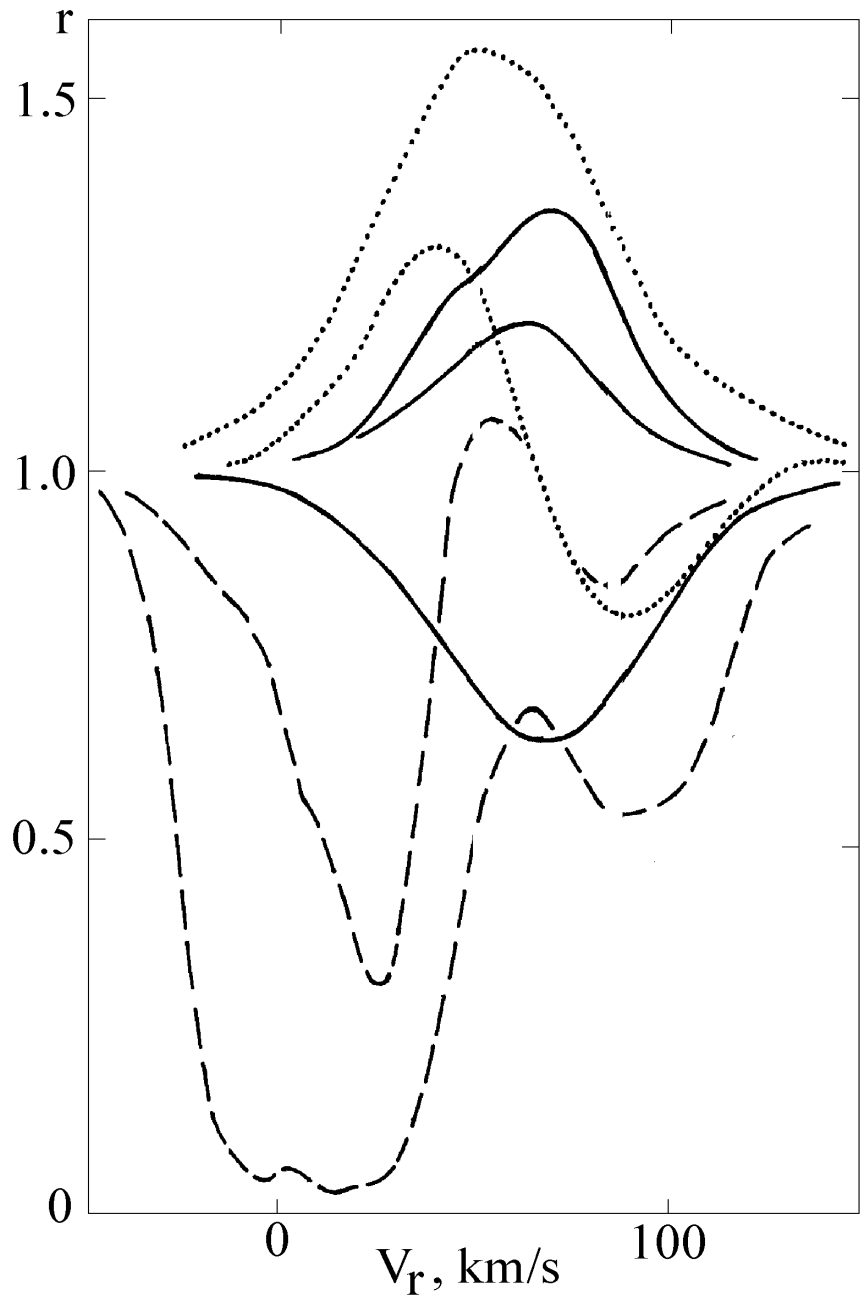,width=14.0cm,height=18.0cm}}
\par
\caption{Line profiles  in the optical spectrum of IRC +10420.
        From top to bottom:
        dotted lines - FeII(40) 6432, FeII(74) 6456;
        solid lines - [FeII] 14F 7155, [OI] 1F 6300, NI(3) 7468;
        dashed lines - KI(1) 7699, NaI(1) 5890.}
\end{figure}

When comparing our radial velocity values with the data of other authors,
aside from the profile variation from line to line and temporal variations
it is necessary to take into account their asymmetry. It is relatively
simple to compare more or less pure emission and absorption lines.
The emission lines (first and third entries of Table\,2) give
${\rm V_r = + (60 - 66)\,km/s}$, which is correlated with
+(61 - 65)\,km/s obtained from the optical, IR and radio emission lines
(Olofsson et al., 1982; Fix and Cobb, 1987; Jones et al., 1993;  Oudmaijer, 1995;
Oudmaijer et al., 1996) and regarded usually as a "systemic velocity".
For the absorption lines the agreement is worse: ours + 66\,km/s for NI(3) and
+72\,km/s for SiII(2) and +74 and +78\,km/s, respectively from Oudmaijer (1995).
The difference is likely to be connected with both different measuring methods
and temporal variability.

For the asymmetric lines, in Table\,2 are presented the radial
velocities obtained from the peaks of the emissions details or from
the cores of the absorption lines, the symbol ${\le}$ and ${\ge}$ in front
of them shows the type of asymmetry.
The type of asymmetry as a rule is the same as in Oudmaijer (1995).
The double-peaked structure of the emission lines ${\rm H_{\alpha}}$ and
${\rm H_{\beta}}$ is likely to be followed in some ion lines as well,
in particular, in the profiles of FeII(74) one can suspect a weak emission
component with ${\rm V_r = +120-130\,km/s}$ apart from the main peak
with ${\rm V_r = +40\,km/s}$.
On the other hand in the absorption lines of FeII (42)
and  part of lines ScII, TiII, CrII the small redshift of the wings with
respect to the cores is apparently due to the presence of a blueward-shifted
emission component.

\begin{figure}
\par
\centerline{\psfig{figure=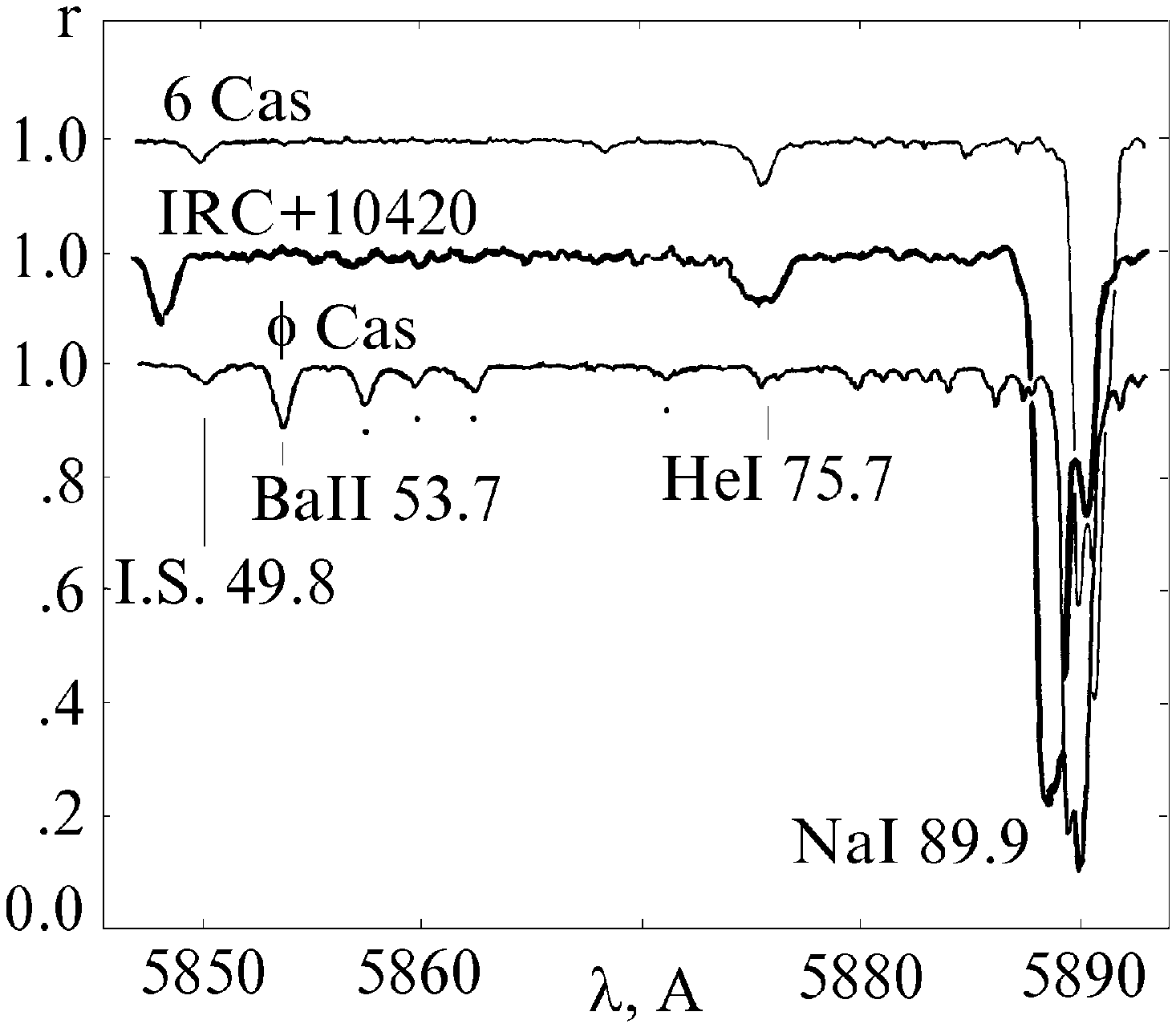,width=17cm,height=14.0cm}}
\par
\caption{Comparison of the IRC +10420 spectrum into the region
         ${\rm 5848 - 5894 \AA}$ with that for supergiants 6 Cas
         A2.5 Ia-0 and ${\rm \phi Cas}$ F0Ia.
         Absorption lines of neutral metals (mainly - FeI) are marked by dots.}
\end{figure}

The relation between velocity and line strength (Oudmaijer, 1995) one can
reveal from our data too.
The "pure" emissions of multiplets 40 and 46 of FeII show the radial velocities
from  +55\,km/s for the strongest line $\lambda $ 6516${\rm \AA}$ to
+64\,km/s for the weakest one $\lambda $ 6116${\rm \AA}$ and
velocities measured for the members of multiplet 73 change
from +40\,km/s to +63\,km/s.

Note also that in the deepest part of the absorption of NaI and KI two
main components with ${\rm V_r \approx -5\,km/s}$ and
${\rm \approx +25\,km/s}$ are recognized.
Both of them, at least the first one, are interstellar completely or to
a considerable degree: the lines NaI and CaII  in the spectra of early
type stars in the vicinity of IRC+10420 show close radial velocities.
In particular, one of those stars is heavily reddened supergiant
HD\,183143~B7Iae.
Interstellar absorptions of KI in its  spectra  consist of two components of
almost equal intensity with the radial velocities about -11 and +4\,km/s
(Chaffee and White, 1982; Herbig and Soderblom, 1982).
By our resolution they correspond to the values in the two last entries
in Table\,2.
The likeness between the blue part of the NaI line profile and profiles
of diffuse interstellar bands in the spectrum of IRC+10420
(both in a shape and velocity) is noticeable.
But it is nessesary to keep in mind that the DIB possess intrinsic
(not kinematic) asymmetry as well.

Can any of the radial velocity values that we have obtained be referred to the
center of mass of the star? Apparently only those that have been measured
 from both almost symmetric emission and absorption lines (entries 1-3  of
Table\,2).
One can consider also different, more complex lines as a natural consequence
of superposition of non-shifted photospheric absorption lines and
 "inverse P\,Cyg profiles" with blue-shifted emission and red-shifted
absorption features.

When considering IRC+10420 as a massive star, it is necessary that its velocity
should be matched with its distance. The galactic longitude of our object is
such (${\rm l = 47^{\circ}}$) that it falls nearly in the center of the
${\rm 30^{\circ}}$  window free from associations and poor in bright
supergiants and O-stars (Humphreys, 1978).
The line of sight passes between the Car-Sgr arm and its local branch (e. g.
see Fig.\,7 in Myers et al. (1986)). Coupled with the strong reddening, this
suggests that the object is very distant. But even at the tangent point (distance
of about 7\,kpc) its heliocentric radial velocity produced by differential
rotation is no higher than 60\,km/s, which is slightly less than the
velocities obtained both from lines with minimal asymmetry
and IR and radio data.

The latter fact together with  "inverse P\,Cyg profiles" of many lines, as well
the character of their temporal variations according to Oudmaijer (1995) (both
emission and absorption components become stronger or grow weak simultaneously)
could be regarded  as evidence of accretion. But it should be noted  also
and some difficulty of this conception.
Indeed,   normal P\,Cyg profiles has been recorded
by Fix and Cobb (1987) and by Oudmaijer (1995) for the IR line of CO,
which is formed in the extended fossil envelope. The blue shift of the
absorption core with respect to the emission peak suggests that the
envelope is expanding at a velocity of 40\,km/s.
Close values, from 40 to 50, are obtained also from the widths of the envelope
emissions in our spectra. It is in good agreement with the data of other
authors obtained in both the optical and the IR, and radio wavelength ranges.
One should not exclude that the lines of NaI\,(1) and KI\,(1) also contain
ordinary P\,Cyg-profiles superimposed on photosperic absorptions and distorted
by the interstellar components.
The values close to 40\,km/s are obtained from the widths of the molecular
and forbidden atomic emission lines.
As it is obvious both from our Fig.\,1 and Oudmaijer's data (1995),
these values for permitted emission lines are even greater.
But the interpretations of the line broadening are opposite: expansion
for the forbidden lines and compression for the permitted ones.
According to such a model the profiles for the forbidden emission lines
are shaped by matter going both up and down consequently by matter immobile
relative to the star.
However, these profiles have not any narrow details to mark the velocity
of the star center of mass. We have no distinct impression that the red
wings of photospheric absorption lines contain the strong envelope
components.
At least, the latter do not prevent from the quantitative spectral
classification.

Humphreys et al. (1973) was the first to estimate the spectral class of
IRC+10420 as F8I. Since then and untill recently (Oudmaijer, 1995;
Oudmaijer et al., 1996) it has apparently never been revised, although
Fix (1981) has noted that the absorption lines in IRC+10420 are weaker
than in ${\rm \gamma\,Cyg}$ F8Ib.
This estimate is one of the signs that  IRC+10420 is similar to
the unique object ${\rm \eta\,Carinae}$, which also showed the spectrum of an
F-supergiant in 1893 (Walborn and Liller, 1977).
Over the last 20 years the expected rise in temperature of IRC+10420
has actually taken place, not as great as in ${\rm \eta\,Carinae}$ though.
Fig.\,2  shows that in 1995-1996 the object IRC+10420 is much closer
to 6\,Cas A2.5 Ia-0 in spectrum than to ${\rm \phi\,Cas}$ F0\,Ia -
the absorptions of the neutral metals are absent, but  the HeI
line ${\rm \lambda 5876 \AA}$ can be seen.
We estimate the spectral class of IRC+10420 using spectral criteria
developed for high-dispersion spectra of normal supergiants and for both blue
and red lines (data by Oudmaijer (1995) and our own, respectively). Our
evalution of spectral class is A5 with the spread from A3 to A7 for
individual criteria.

The rise in effective temperature in IRC+10420 allows us to more seriously
treat its likeness to, possibly, even a relationship with B[e] supergiants.
In the last time the domains of this group has been extended in luminosity
from ${\rm 10^{6}}$ to ${\rm 10^{4}\,L_{\odot}}$ and to B9 in spectral class.
The main thing is that spectra have been obtained in which the absorptions
show small, occasionally even red shifts with respect to the emission
features (Zickgraf et al., 1992; Gummersbach et al., 1995).
Moreover, on the (J-H)-(H-K) diagram IRC+10420 falls within the same
isolated region that is occupied by B[e] stars due to their IR excess.

\section{Estimation of metallicity}

For understanding  of an object at an unclear evolutionary stage,
it is very important to know its metallicity and chemical abundance pattern.

To study in detail the chemical composition of an unstable supergiant
with the presence of strong  gradient of radiative pressure and velocity field,
one can not use classical model atmospheres method. One needs  dynamic
inhomogeneous spherical models in this case.
Therefore it should be noted that we present here the preliminary results only
on chemical composition  of IRC+10420 because we use the standard
plane-parallel homogeneous models grid of Kurucz (1979).

For chemical composition calculation by the model atmosphere method, one needs
to know the values of the effective temperature ${\rm T_{eff}}$, surface gravity
log\,g and microturbulent velocity ${\rm \xi_t}$.
The most difficult problem is effective temperature determination.
Determination of ${\rm T_{eff}}$ is problematic even for normal supergiants
due to their extended atmospheres and significant non-LTE effects
(Venn, 1993, 1995a,b).
The problem arises for such a peculiar star as IRC+10420, for which the energy
distribution is strongly distorted by interstellar and circumstellar extinction.
We can not use also in the case of IRC+10420 the equivalent widths and profiles
of hydrogen lines for ${\rm T_{eff}}$ and log\,g determination.
These spectral features being well known criteria of atmospheric
conditions for the atmospheres of normal A-supergiants are hardly
distorted in the spectrum investigated.

\tablecaption{Atomic data, equivalent widths W of lines in the
              spectrum of IRC+10420}
\tabletail{\rule{0pt}{1pt}&&&&&\\ \hline}
\begin{supertabular}{llrrrr}
$\lambda, \AA $ & Species & EP, ev & log\,gf & ${\rm W, m\AA}$&${\rm \epsilon(X)}$ \\
\hline
 5172.70 &    Mg   &      2.71 &  -.38  &  184  &     7.42   \\
 4128.05 &    Si+  &      9.84 &   .57  &  237  &     7.05   \\
 5978.93 &    Si+  &     10.07 &  -.06  &  101  &     7.16   \\
 4325.01 &    Sc+  &       .60 &  -.44  &  230  &     3.51   \\
 4400.36 &    Sc+  &       .61 &  -.63  &  226  &     3.69   \\
 5239.81 &    Sc+  &      1.45 &  -.77  &   53  &     3.51   \\
 5526.79 &    Sc+  &      1.77 &   .13  &  180  &     3.47   \\
 5640.99 &    Sc+  &      1.50 & -1.01  &   30  &     3.49   \\
 5657.91 &    Sc+  &      1.51 &  -.50  &  178  &     3.90   \\
 5667.15 &    Sc+  &      1.50 & -1.20  &   15  &     3.36   \\
 5669.04 &    Sc+  &      1.50 & -1.09  &   38  &     3.68   \\
 5684.20 &    Sc+  &      1.51 & -1.01  &   22  &     3.35   \\
 4028.33 &    Ti+  &      1.89 &  -1.12 &  169  &     5.00   \\
 4287.89 &    Ti+  &      1.08 &  -1.68 &  185  &     5.02   \\
 4316.81 &    Ti+  &      2.05 &  -1.52 &   57  &     4.87   \\
 4386.86 &    Ti+  &      2.60 &  -1.11 &  141  &     5.31   \\
 4411.08 &    Ti+  &      3.09 &   -.82 &  131  &     5.31   \\
 4411.94 &    Ti+  &      1.22 &  -2.32 &   66  &     5.17   \\
 4421.95 &    Ti+  &      2.06 &  -1.43 &  112  &     5.13   \\
 4464.46 &    Ti+  &      1.16 &  -1.78 &  180  &     5.14   \\
 4470.86 &    Ti+  &      1.16 & -2.00  &  119  &     5.11   \\
 4488.32 &    Ti+  &      3.12 &  -.86  &  156  &     5.47   \\
 4529.47 &    Ti+  &      1.57 & -1.93  &   59  &     4.95   \\
 4544.01 &    Ti+  &      1.24 & -2.28  &   43  &     4.92   \\
 4568.31 &    Ti+  &      1.22 & -2.52  &   32  &     5.01   \\
 4779.98 &    Ti+  &      2.05 & -1.37  &  125  &     5.09   \\
 4805.10 &    Ti+  &      2.06 & -1.05  &  198  &     5.06   \\
 4874.01 &    Ti+  &      3.09 &  -.79  &  120  &     5.19   \\
 4911.18 &    Ti+  &      3.12 &  -.34  &  190  &     5.05   \\
 5129.16 &    Ti+  &      1.89 & -1.39  &  192  &     5.24   \\
 5185.90 &    Ti+  &      1.89 & -1.35  &  170  &     5.12   \\
 5211.53 &    Ti+  &      2.59 & -1.85  &   40  &     5.33   \\
 5336.78 &    Ti+  &      1.58 & -1.70  &  114  &     5.00   \\
 5381.01 &    Ti+  &      1.57 & -2.08  &   76  &     5.16   \\
 3951.97 &    V +  &      1.48 &  -.52  &  235  &     4.16   \\
 4023.39 &    V +  &      1.80 &  -.72  &   95  &     3.99   \\
 4600.19 &    V +  &      2.27 & -1.31  &   16  &     3.98   \\
 4252.62 &    Cr+  &      3.86 & -2.10  &   78  &     5.78   \\
 4261.92 &    Cr+  &      3.86 & -1.73  &  154  &     5.81   \\
 4269.28 &    Cr+  &      3.85 & -2.33  &   87  &     6.06   \\
 4275.57 &    Cr+  &      3.86 & -1.85  &  186  &     6.05   \\
 4284.21 &    Cr+  &      3.85 & -1.91  &  105  &     5.75   \\
 4555.02 &    Cr+  &      4.07 & -1.45  &  228  &     5.93   \\
 4592.07 &    Cr+  &      4.07 & -1.35  &  239  &     5.88   \\
 4616.64 &    Cr+  &      4.07 & -1.35  &  173  &     5.62   \\
 4812.35 &    Cr+  &      3.86 & -1.89  &   67  &     5.46   \\
 5279.88 &    Cr+  &      4.07 & -2.10  &   37  &     5.50   \\
 5308.46 &    Cr+  &      4.07 & -1.81  &   48  &     5.34   \\
 5313.61 &    Cr+  &      4.07 & -1.65  &  120  &     5.66   \\
 5334.88 &    Cr+  &      4.07 & -1.89  &  126  &     5.92   \\
 4045.82 &    Fe   &      1.48 &   .28  &  215  &     7.37   \\
 4063.60 &    Fe   &      1.56 &   .06  &   93  &     7.11   \\
 4071.74 &    Fe   &      1.61 &  -.01  &  107  &     7.29   \\
 4132.06 &    Fe   &      1.61 &  -.63  &   43  &     7.44   \\
 4181.76 &    Fe   &      2.83 &  -.31  &   24  &     7.67   \\
 4210.35 &    Fe   &      2.48 &  -.95  &   15  &     7.86   \\
 4250.13 &    Fe   &      2.47 &  -.41  &   10  &     7.12   \\
 4250.79 &    Fe   &      1.56 &  -.73  &   63  &     7.68   \\
 4271.16 &    Fe   &      2.45 &  -.35  &   16  &     7.26   \\
 4271.76 &    Fe   &      1.48 &  -.16  &   92  &     7.25   \\
 4325.76 &    Fe   &      1.61 &  -.02  &  175  &     7.57   \\
 4383.55 &    Fe   &      1.48 &   .20  &  222  &     7.44   \\
 4404.75 &    Fe   &      1.56 &  -.14  &  144  &     7.53   \\
 4122.67 &    Fe+  &      2.58 & -3.62  &  158  &     7.47   \\
 4128.74 &    Fe+  &      2.58 & -4.00  &  100  &     7.56   \\
 4273.32 &    Fe+  &      2.70 & -3.53  &  123  &     7.28   \\
 4541.52 &    Fe+  &      2.86 & -3.21  &  218  &     7.44   \\
 4582.84 &    Fe+  &      2.84 & -3.44  &  131  &     7.30   \\
 4620.51 &    Fe+  &      2.83 & -3.61  &   78  &     7.16   \\
 4666.75 &    Fe+  &      2.83 & -3.57  &  117  &     7.35   \\
 6247.55 &    Fe+  &      3.89 & -2.51  &  192  &     7.30   \\
 6456.39 &    Fe+  &      3.90 & -2.30  &  234  &     7.27   \\
 4077.71 &    Sr+  &      0.00 & 0.21   &  253  &     2.63   \\
 4215.52 &    Sr+  &      0.00 & 0.04   &  140  &     2.36   \\
 4177.54 &    Y +  &      0.40 & -0.24  &  177  &     3.28   \\
 4398.02 &    Y +  &       .12 &  -.14  &   52  &     2.28   \\
 4149.20 &    Zr+  &       .79 &  -.13  &   85  &     2.97   \\
\end{supertabular}

We can use the spectroscopic method of temperature determination only, forcing
independence of the abundance derived for each line upon the excitation
potential of the low level for this line. The parameter log\,g was estimated through ionization
balance for FeI and FeII.
The microturbulent velocity value based on W of CrII, FeII, ScII and TiII lines
is very high and equal to 12 km/s. But such a high, close to the velocity of sound,
value of microturbulent velocity is trivial for the most luminous hot supergiants
at the limit of the HR-diagram (Nieuwenhuijzen, de Jager, 1992).

It is well known that the plane-parallel static model atmosphere method does not
give correct abundances for high luminosity stars, (Ia, Ia+).
The profiles of the spectral lines observed are additionally broadened by
non-thermal mecha\-nisms whose influence may be variable at different levels
in the atmosphere.
To obtain more reliable estimates of abundances we use only weak lines with
${\rm W < 250\,m\AA}$.
Such lines formed closer to photospheric layers are more
correctly described by a standard stationary model.

All lines which have been used for chemical composition calculation are listed
in Table\,3. Here are given also equivalent widths W, oscillator strengths
log\,gf used, excitation potentials EP of lower level and abundances calculated
for individual lines.
It should be noted that for abundance calculations we used equivalent widths
derived from both our spectra of IRC+10420 and the spectrum kindly provided
by R.Oudmaijer.
For most lines our W coincided with that measured by Oudmaijer (1995).

\begin{figure}
\par
\centerline{\psfig{figure=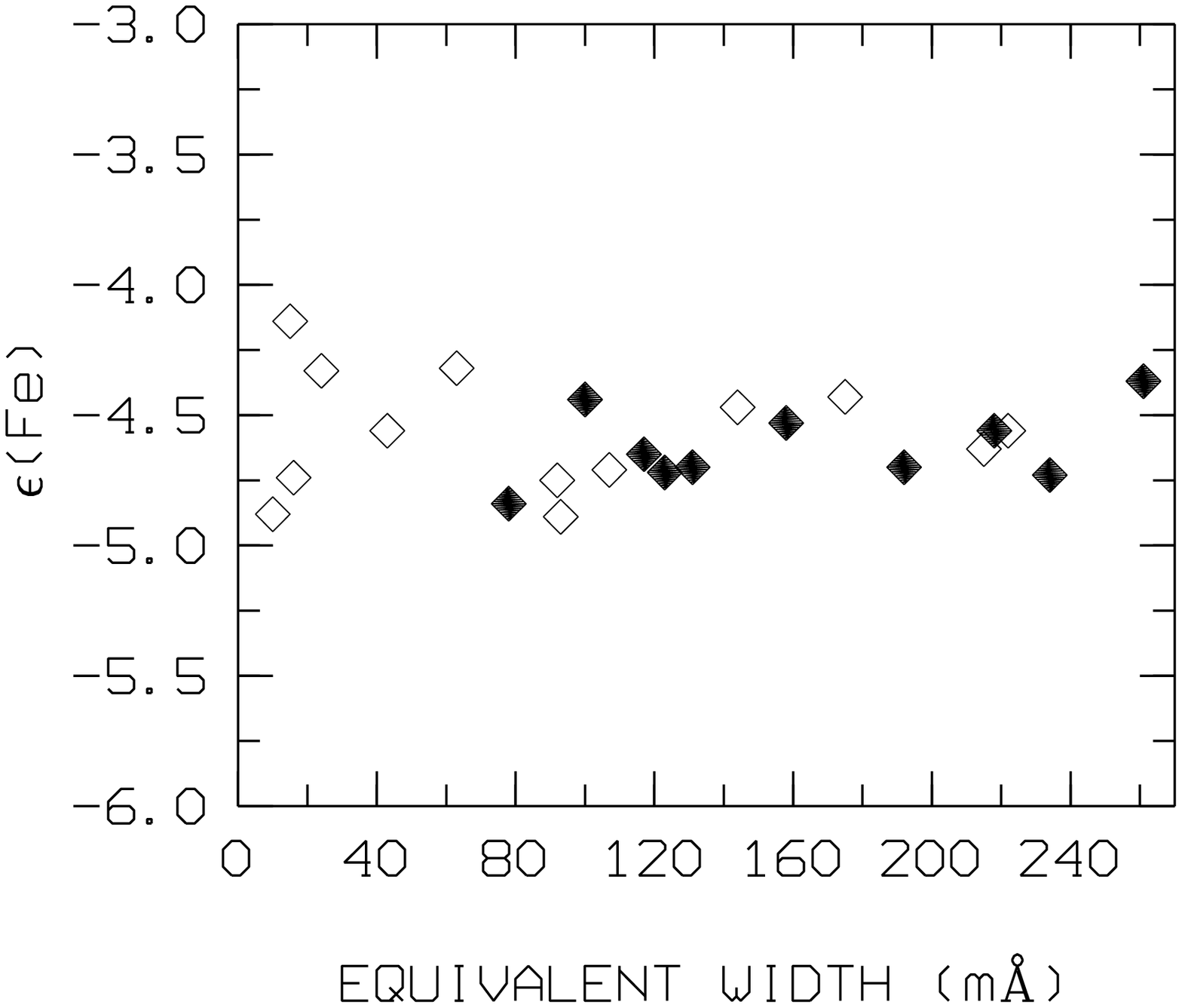,width=14cm,height=10.0cm}}
\par
\centerline{\psfig{figure=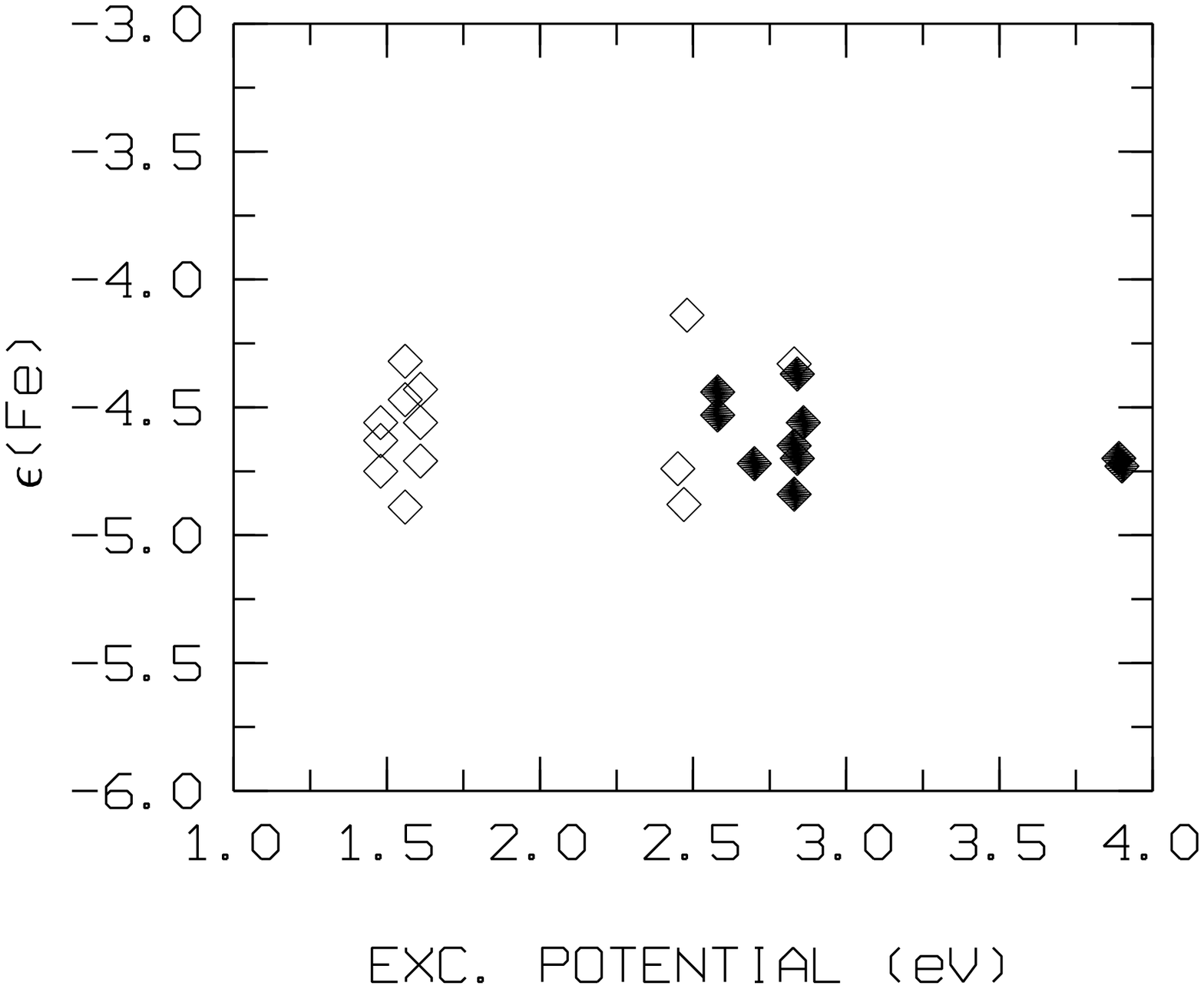,width=14cm,height=10.0cm}}
\caption{Iron abundance derived from individual FeI (open symbols) and FeII
(filled symbols) lines as a function of line equivalent width W (upper panel)
and as a function of excitation potential (lower panel).}
\end{figure}

The model atmosphere parameres and abundances of several chemical elements
obtained are presented in Table\,4.
The abundances are given on the scale
${\rm \epsilon\,(X) = log\,N(X)-log\,N(H)}$,  at ${\rm log\,N(H)=12.0}$.

The value for ${\rm T_{eff} = 8500\,K}$ derived from the intensity of neutral
iron absorption lines is well fit for the value of spectral class
${\rm A5\pm 0.5}$ determined from the W of a sample of lines of different
chemical elements (FeII, CrII, TiII).
The internal uncertainty of the main model parameter ${\rm T_{eff}}$ is
equal to ${\rm \Delta T_{eff} = \pm 250\,K}$, which corresponds to the
accuracy of spectral classification.
The error of the log\,g value derived from ionization
equilibrium of iron is determined by forcing a maximum difference between
${\rm \epsilon\,(FeI)}$ and ${\rm \epsilon\,(FeII)}$ of 0.1\,dex. Such
difference is achieved by varying of log\,g by ${\rm \pm 0.2}$ at other
constant parameters (${\rm T_{eff}}$ and ${\rm \xi_t}$).
The value of ${\rm \xi_t}$ is determined with an uncertainty of
${\rm \pm 1.0\,km/s}$, which is typical for A-supergiants (Venn, 1993).
The calculated abundance errors caused by uncertainties of the atmospheric
parameters ${\rm T_{eff}}$,  log\,g and  ${\rm \xi_t}$ are given in Table\,5.
As it shown in this Table, the limitation of equivalent width of lines used
${\rm W < 250\,m\AA}$ reduces significantly the influence of uncertainty of
${\rm \xi_t}$ choice. The main factor of abundance errors for most species is
the uncertainty of   ${\rm T_{eff}}$ value.

As it is shown in Figure\,3, there is no relationship between
${\rm \epsilon(Fe)}$ and equivalent widths W and excitation
potentials EP of iron lines which  have been used for metallicity
estimation of IRC+10420. It illustrates sufficiently correct choice
of model parameters.

A lot of absorption lines of ionized metals (Ti, Sc, Cr) have been reliably
measured in the spectrum of IRC+10420. It is important that we have not found
any dependence of abundance of these ions neither on W or EP. Therefore the
microturbulent velocity does not vary between different chemical elements.

As it is shown in Table\,4, abundances of a number of species (ScII, TiII,
VII, CrII, FeI, FeII) are determined with a high internal accuracy,
${\pm \sigma \le 0.03 - 0.06}$.
It allows us to conclude that with a high probability the metallicity
for IRC+10420 is close to  solar: the average abundance for elements of
iron-group relative to the Sun is ${\rm \epsilon(V,Cr,Fe)=-0.03}$.

\begin{table}
\caption{Model atmosphere parameters adopted and abundances of chemical
         elements for IRC+10420. ${\rm log\,N(H)=12.0}$. n - number of
         lines used for calculation. The data by Grevesse and Noels (1993) are
         adopted for solar abundances.}
\begin{tabular}{lccrc}
&&&& \\
\hline
\multicolumn{5}{c}{${\rm T_{eff} = 8500\,K, log\,g=1.0, \xi_t = 12\,km/s}$} \\[5pt]
\hline
Element&${\rm \epsilon (X)}$ & ${\pm \sigma}$ & n & ${\rm [X/H]_{\odot}}$\\[5pt]
\hline
 MgI          &  7.42 &       &  1 & -0.12  \\
 SiII         &  7.10 & 0.06  &  2 & -0.45  \\
 ScII         &  3.55 & 0.06  &  9 & +0.38  \\
 TiII         &  5.12 & 0.03  & 22 & +0.10  \\
 VII          &  4.04 & 0.06  &  3 & +0.04  \\
 CrII         &  5.76 & 0.06  & 13 & +0.09  \\
 FeI          &  7.43 & 0.06  & 13 & -0.07  \\
 FeII         &  7.35 & 0.04  &  9 & -0.15  \\
 SrII         &  2.50 &       &  2 & -0.40  \\
 YII          &  2.78 &       &  2 & +0.54  \\
 ZrII         &  2.97 &       &  1 & +0.37  \\
\hline
\end{tabular}
\end{table}

The $\alpha$-process element silicon is underabundant, at the
same time the ScII abundance well determined from 9 lines is increased
relative to Fe: ${\rm [Sc/Fe]_{\odot} = +0.49 dex}$.
The derived underabundance of silicon may be real and caused by the
selective depletion of a part of this chemical element due to condensation
to silicate dust in the circumstellar envelope and following accretion
of such "cleaned" gas onto the photosphere.
Similar processes are known for stars with IR-excesses: $\lambda$\,Bootis
stars (Venn, Lambert, 1990; St\"urenburg, 1993) and post-AGB stars
(van Winckel, 1997).
In connection with the silicon photospheric underabundance it should
be noted that the spectral energy distribution (SED) of IRC+10420 fits
to a two-component shell model, and SED for the inner hotter shell fits
to {\bf silicate} dust (Oudmaijer et al., 1996).

\begin{table}
\caption{Estimated errors of chemical elements abundances due to uncertainties
of the atmospheric parameters ${\rm T_{eff}}$, log\,g and ${\rm \xi_t}$.}
\begin{tabular}{lccc}
&&& \\
\hline
Element & \multicolumn{3}{c}{${\rm \Delta(X)}$} \\[5pt]
\hline
 &${\rm \Delta T_{eff} = -250\,K}$ &${\rm \Delta log\,g = +0.2}$ &
${\rm \Delta \xi_t = -1\,km/s}$\\[5pt]
\hline
 MgI          &  -0.50 &  -0.25 & 0.03  \\
 SiII         &  +0.07 &  +0.10 & 0.06  \\
 ScII         &  -0.46 &  -0.15 & 0.01  \\
 TiII         &  -0.34 &  -0.10 & 0.01  \\
 VII          &  -0.29 &  -0.07 & 0.02  \\
 CrII         &  -0.18 &  -0.00 & 0.04  \\
 FeI          &  -0.51 &  -0.26 & 0.01  \\
 FeII         &  -0.17 &  -0.00 & 0.02  \\
\hline
\end{tabular}
\end{table}

Individual abundances of the heavy s-process metals Sr, Y, Zr are determined
with a large error because of the small number of lines measured.
But the average abundance close to the solar value of [X/Fe] for
these s-process metals is sufficiently reliable.

All these details of chemical composition for IRC+10420 is similar to the average
chemical composition of massive A-supergiants (Venn, 1995a).

Unfortunately,  within the frames of standard models, we can not derive
quantative conclusions concerning nitrogen and oxygen content
for IRC+10420 because in its spectrum only very strong lines of these elements
are observed, with ${\rm W > 300-400\,m\AA}$.
To obtain reliable abundances of these elements the LTE-approach is
insufficient  for the case of A-supergiants (Venn, 1995b).

But we can draw qualitative  conlusions about the  behaviour of C, N-lines
in the IRC+10420 spectrum from a comparison of their equivalent widths  with ones
for normal supergiants of  similar effective temperature and surface gravity.
We have used for the comparison the equivalent widths of carbon and nitrogen lines
in the spectrum of the massive supergiant HD\,13476 (A3Iab) for which Venn
(1995a) has derived atmospheric parameters:
${\rm T_{eff} = 8400\,K}$, ${\rm log\,g=1.2}$, ${\rm \xi_t = 8\,km/s}$.
Abundances of some metals for HD\,13476 are solar and very close to
that of IRC+10420 estimated here.
Later Venn (1995b) calculated non-LTE C-, N-abundances for 22 A-type
supergiants including HD\,13476.
Carbon and nitrogen relative to the iron content for HD\,13476
${\rm [C/Fe]_{\odot}]=-0.43}$, ${\rm [N/Fe]_{\odot}]=0.08}$ are close to
the averages for the whole sample studied. The non-LTE corrections are vary:
${\rm \Delta = -0.42}$ for carbon and ${\rm \Delta = -0.80}$ for nitrogen.
The value ${\rm [N/C]_{\odot} = 0.51}$ is very similar to the first dredge-up
for stars with mass near ${\rm 8-10\,M_{\odot}}$.

The equivalent widths of N lines in the spectrum of IRC+10410 are much
stronger (2-3 times) in comparison to the W-values of the same lines
in the HD\,13476 spectrum and the equivalent widths of C lines are essentially
weaker (see Table\,6).
Therefore we can conclude that ${\rm [N/C]_{\odot}}$--value for IRC+10420
in any case is not less than that  for HD\,13476.

\begin{table}
\caption{Comparison of equivalent widths W of C,N-lines for two high luminous
stars HD\,13476 (Venn, 1995b) and IRC+10420 (Oudmaijer, 1995)}
\begin{tabular}{ccc}
&& \\
\hline
$\lambda,\AA$ & \multicolumn{2}{c}{${\rm W, m\AA}$} \\[5pt]
\cline{2-3}
&  HD\,13476 & IRC+10420 \\
\hline
CI: &&\\
9088.57 &240&67\\
9094.89 &392&260\\
9111.85 &240&68\\
NI:&&\\
7423.63 &95&288\\
7442.28 &164&434\\
7468.29 &214&604\\
8184.80 &230&491\\
8187.95 &251&518\\
8210.64 &128&293\\
8216.28 &358&779\\
8242.34 &216&523\\
8629.24 &269&312\\
8703.24 &279&621\\
8711.69 &306&590\\
8718.82 &258&544\\
8728.88 &91 &211\\
\hline
\end{tabular}
\end{table}

\section{Conclusions}

So, the optical spectrum of IRC+10420 of the years from 1992 through 1996
\begin{itemize}
\item{points to the increase in the temperature: spectral class A5 instead
    of  F8 in 1973;}
\item{contains absorptions (mainly of ions) formed in the photosphere,
      apparently stationary with respect to the star center of mass;}
\item{contains emission details too, which are formed in the expanding envelope
      and perhaps in its compressing layers;}
\item{rezembles the spectra of late-type B[e] stars;}
\end{itemize}

The metallicity,  which is close to solar, and the altered [N/C]-value
allow us to consider
IRC+10420 as a massive object at the evolution stage of at least after the first
dredge-up and  not to reject the hypothesis of Jones et al. (1993) who suggested
that IRC+10420 is a true hypergiant with a mass of about
${\rm 40\,M_{\odot}}$, evolving from the red supergiant phase to
the Wolf-Rayet stage.

The nature of IRC+10420 is still open question. New observations are
necessary to obtain full chemical abundance pattern and to coordinate luminosity,
distance and radial velocity of this object. Besides that we need more careful
indentification of extremely plentiful and variable features of the optical
spectra of IRC+10420. For example, what are the enigmatic narrow strong
absorption details ${\rm \lambda\lambda\,5694,6286, 6288\,\AA}$ ? For the
present one can say only that these are not telluric lines.

It should be noted that in the frames of the study  of supergiants with
infrared excess (Klochkova, 1995; Za\v{c}s et al., 1995) we have obtained
also the  spectra of the supergiant HD179821 = IRAS19114+0002, which has
been considered analogous to IRC+10420 (Kastner, Weintraub, 1995).
Based on these spectra results have recently been  obtained
(Za\v{c}s et al., 1996) for HD179821, using the model atmosphere
method: ${\rm T_{eff} = 6800\,K}$, log\,g=1.3, iron peak elements are
slightly underabundant, Sc and Ti are slightly overabundant.
But for this object Za\v{c}s et al. (1996) revealed  the overabundance of
sodium, s-process (Y, Zr) and r-process (Eu) elements.
The overabundance of s-process points to the post-AGB stage of evolution of
HD\,179821. However for more general conclusions about the nature of HD\,179821
and its likeness to IRC+10420, abundances of CNO-triad are needed.

{\it Acknowledgements.} We thank Prof. Yu.\,N.\,Efremov for helpful
discussion on localization of IRC+10420 in the Galaxy and Dr. R.\,Oudmaijer
for a part of his Ph.D.\,Thesis kindly made available to us.
We are also indebted  to the referee for detailed and fruitful discussion and
suggestions.

Program of spectroscopic research of IRAS-sources has financial support from
the Russian Federal Program "Astronomy". We acknowledge also the partial
support from KBN grant No.2P03D.026.09.

}

\end{document}